\documentclass[10pt, journal]{IEEEtran}
\usepackage{amsfonts}
\usepackage{amssymb}
\usepackage{algorithm}
\usepackage{algpseudocode}
\usepackage{mathtools}  
\usepackage{amsthm}
\usepackage{subfigure}
\usepackage{tabulary}
\usepackage{footnote}
\usepackage[export]{adjustbox}
\usepackage{booktabs}
\usepackage{multirow}
\usepackage{comment}

\usepackage{cases}
\usepackage{cite}
\usepackage{multicol}
\usepackage{threeparttable}
\usepackage{xcolor}
\usepackage{graphicx}
\usepackage{hyperref}
\graphicspath{{./}{figures/}}

\usepackage{stfloats}
\usepackage{cuted}
\usepackage{lipsum}

\usepackage{float}
\usepackage{makecell}
\usepackage{caption} 
\usepackage{graphicx}

\title{A Lightweight Coordinate-Conditioned Diffusion Approach for 6G C-V2X Radio Environment Maps}

 \author{ \text{Liu Cao}\IEEEauthorrefmark{1}, \text{Zhaoyu Liu}\IEEEauthorrefmark{1}, \text{Dongyu Wei}, \text{Yuan Yang}, \text{Yukun Pan}, \text{Lyutianyang Zhang}
\vspace{-0.5cm}

 \thanks{\IEEEauthorrefmark{1}Both authors contributed equally to this work. Liu Cao, Zhaoyu Liu, Yuan Yang, and Yukun Pan are with both City University of Hong Kong (Dongguan), Dongguan, China, and City University of Hong Kong, Hong Kong (e-mail:\{liu.cao,72515198,72515475,72515531\}@cityu-dg.edu.cn). Dongyu Wei is with the Department of Electrical and Computer Engineering, University of Miami, Coral Gables, FL, USA (e-mail: dongyu.wei@miami.edu). Lyutianyang Zhang is  with
the School of Microelectronics and Communication Engineering, Chongqing
University, Chongqing, China (email: zhanglyutianyang@cqu.edu.cn).  (Corresponding author: Lyutianyang Zhang.)}

\thanks{This work was supported by the Youth Innovation Talent Project of Guangdong Provincial Universities (Grant No. 2025KQNCX17).}
}

\begin{document}

\maketitle
\begin{abstract}
Transmitter vehicles that broadcast 6G Cellular Vehicle-to-Everything (C-V2X)-based messages, e.g., Basic Safety Messages (BSMs), are prone to be impacted by PHY issues due to the lack of dynamic high-fidelity Radio Environment Map (REM) with dynamic location variation. This paper explores a lightweight diffusion-based generative approach, the Coordinate-Conditioned  Denoising Diffusion Probabilistic Model (CCDDPM), that leverages the signal intensity-based 6G V2X Radio Environment Map (REM) from limited historical transmitter vehicles in a specific region, to predict the REMs for a transmitter vehicle with arbitrary coordinates across the same region. The transmitter vehicle coordinate is encoded as a smooth Gaussian prior and fused with the Gaussian noise through a lightweight two-channel conditional U-Net architecture. We demonstrate that the predicted REM closely matches the statistics and structure of ground-truth REM while exhibiting the improved stability and over other widely applied generative AI approaches. The resulting predictor enables rapid and scenario-consistent REM with arbitrary transmitter coordinates, which thereby supports more efficient 6G C-V2X communications where transmitter vehicles are less likely to suffer from the PHY issues.

\end{abstract}
\begin{IEEEkeywords}
 Generative AI, Diffusion, 6G V2X, Radio Environment Map, Signal Intensity Prediction
\end{IEEEkeywords}

\section{Introduction}
\label{sec:intro}

A high-fidelity Radio Environment Map (REM) is becoming a cornerstone for 6G vehicular networks (V2X) \cite{yapar2022dataset}. An REM provides spatially dense signal intensity fields (e.g., RSRP/RSSI) that downstream modules (handover, relay/forwarding, resource scheduling) can query in real time, avoiding costly ray tracing or repeated drive tests. Maintaining the accurate REM for each vehicle in 6G C-V2X is challenging because vehicles function as moving transmitters with continuously changing poses, urban obstacles induce strong location-dependent multipath, and privacy constraints limit dense centralized data collection; consequently, even within the road segment, small transmitter pose shifts can cause large power changes and yield pronounced Non-IID behavior that undermines generalization from seen to unseen coordinates while offering no uncertainty to guide new measurements.

Recent studies on coverage and signal intensity prediction for vehicular networks mainly adopt BS/RSU-centric radio-environment mapping with supervised regressors trained in centralized pipelines, which limits adaptability to rapidly changing traffic scenes~\cite{10440080}. Federated LSTM methods remain largely focused on indoor REMs or infrastructure-fixed setups, offering limited support for scene-consistent prediction when the transmitter is a moving vehicle within the same road segment~\cite{10437765}. Traditional C-V2X efforts also rely on V2I measurements under fixed RSU layouts to learn RSRP from environmental factors, which is useful for infrastructure planning but not for forecasting spatial fields for moving transmitters or handling time-varying transmit power. In parallel, ISAC-aided 5G/NR-V2X research employs deep reinforcement learning to dynamically tune MCS/SCS/Tx-power, treating power as a control variable for resource management rather than an input-conditioned spatial field to be predicted~\cite{10333693,10554663,10898772,luo2025denoising,cao2022resource}. Motivated by AI-native/semantic communications, we instead target multi-road scenes with moving transmitters and learn a coordinate-conditioned generative mapping from arbitrary transmitter coordinates to dense signal-intensity fields via a Coordinate-Conditioned Denoising Diffusion Probabilistic Models (CCDDPM), enabling per-scene synthesis of C-V2X REM.


This paper considers a complex urban scenario and aim to predict the REMs for arbitrary transmitter vehicle coordinates based on the REMs with limited historical transmitter vehicles across the same region. Our objective is to model the scene-consistent distribution so that we can synthesize high-fidelity REMs at unseen coordinates within the same scene and quantify predictive uncertainty for active updating of the digital twin. The main contributions are summarized as follows:

\begin{itemize}
\item \textbf{Coordinate-Conditioned diffusion framework for 6G C-V2X intensity prediction.}
We propose conditional diffusion models for large-scale non-IID road REM, replacing static, poorly-generalizing methods. Our framework,which uses Gaussian heatmap based on coordinates for  REM encoding,enables robust prediction at unseen locations.
\item \textbf{Low-Overhead Update Paradigm via Sensing and Prediction.}
We integrate generative uncertainty with active updates, maintaining freshness via model baselines while targeting only high-uncertainty areas, thereby slashing measurement costs.
\item \textbf{Edge-Optimized Lightweight Architecture.}
Our conditional diffusion framework incorporates lightweight U-Net with dual-channel diffusion integrates attention and tiling mechanisms. Compatible with few-step samplers, it enables fast, large-scale edge deployment.
\end{itemize}

The rest of this paper is organized as follows: Sec \ref{sec:sys_arc} details the mathematical formulation, model architecture, and presents the proposed CCDDPM. The experimental design used to rigorously evaluate Coordinate-Conditioned DDPM, including the dataset, heterogeneity scenario construction, evaluation metrics, and baseline methods are detailed in Sec \ref{sec:sim}. Finally, section \ref{sec:con} draws the conclusions for this paper.

    \begin{figure*}[t]
        \centering
        \includegraphics[width=0.95\linewidth]{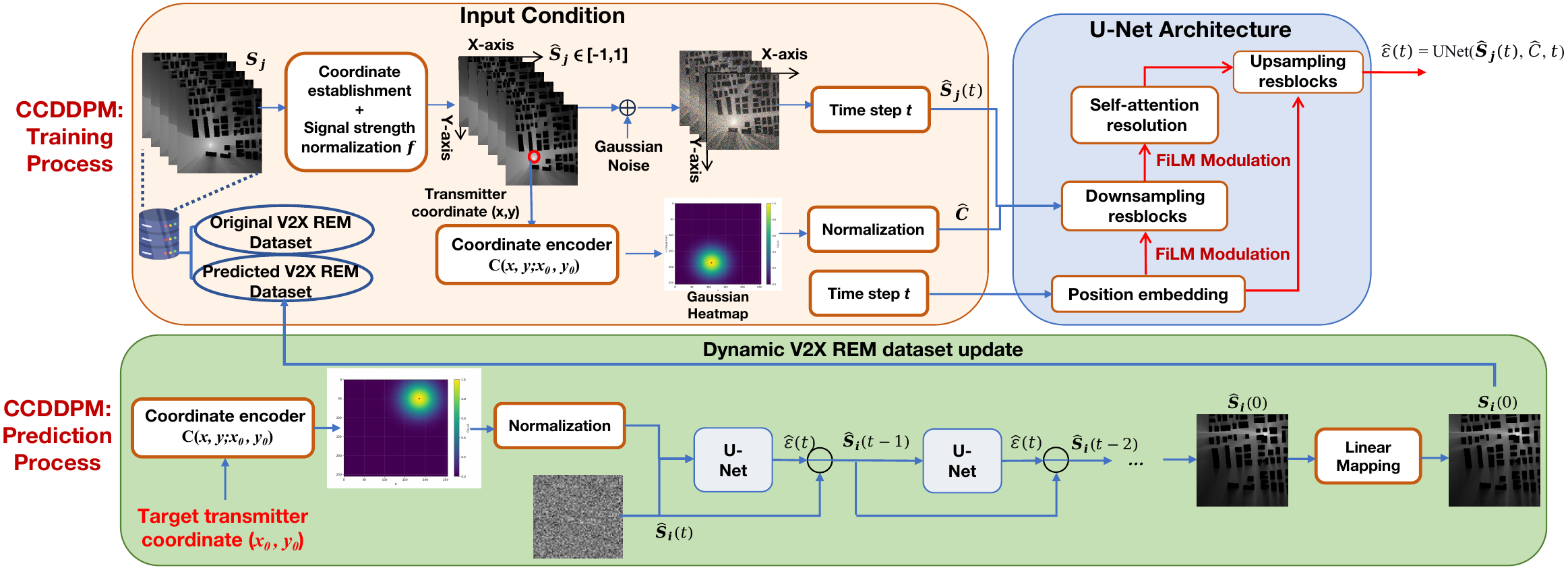}
        \caption{CCDDPM: Training and Prediction Process.}  
        \vspace{-0.5cm}
        \label{fig:P_map}
    \end{figure*}

\section{The Lightweight Conditional Diffusion Approach}
\label{sec:sys_arc}

\subsection{System Architecture}
Let us consider an urban V2X scenario where the wireless propagation over a road network is represented by a radio environment map (REM), where the brightest node corresponds to signal strength of the transmitter vehicle location, and brighter regions indicate stronger received power, while dimmer regions indicate weaker coverage due to distance-dependent path loss and blockage. Black regions denote buildings, which attenuate or block the signal and therefore create shadowed low-power areas behind them. The REM of all transmitter vehicles are collected as the training datasets \footnote{The dataset used in this paper can be found at https://ieee-dataport.org/documents/dataset-pathloss-and-toa-radio-maps-localization-application. }. Our goal is to predict the REM for a transmitter vehicle at any location across the same urban area.  To this end, we use a Cartesian coordinate-conditioned diffusion model that learns the mapping from the transmitter coordinates and environment features to a 2D signal-strength map. The training dataset collected from historical transmitter vehicles across this urban region is
\begin{equation}
\mathcal{D}=\{(\mathbf{u}_j,\mathbf{S}_j)\}_{j=1}^{\mathcal{D}},
\label{eq:trainset}
\end{equation}
where $\mathbf{S}_j\in\mathbb{R}^{H\times W}$ is the REM of the $j$th transmitter vehicle, $H$ and $W$ represent the size (in meters) of the area in the X and Y direction, respectively.  $\mathbf{u}_j=[x,y,\text{env}]\in\mathbb{R}^{2+P}$ combines the coordinate  of transmitter vehicle $(x_0,y_0)$ with $P$ environment features (e.g., diffraction, scattering, reflection). Then we normalize the signal strength values $S_{j}$ into
\begin{equation}
\widehat{\mathbf{S}}_j= f(\widehat{\mathbf{S}}_j)= \frac{2(\mathbf{S}_j-{\mathbf{S}_j}^{min})}{{\mathbf{S}_j}^{max}-{\mathbf{S}_j}^{min}}-1.
\label{eq:pixnorm}
\end{equation}
Thus \(\widehat{\mathbf{S}}_j\in[-1,1]\).  We establish the Cartesian coordinate for the map where the origin (0,0) is assigned at the top-left corner. To provide a smooth spatial prior, the transmitter location $(x_0,y_0)$ is converted into a continuous Gaussian heatmap
\begin{equation}\label{eq:gauss}
\begin{aligned}
& \mathbf{C}(x,y;x_0,y_0)
 =\exp\!\Big(-\tfrac{(x-x_0)^2+(y-y_0)^2}{2\sigma^2}\Big),
\end{aligned}
\end{equation}
where $x \in[0,W\!-\!1],\ y\in[0,H\!-\!1]$, and $(x,y)$ indexes a power location and $\sigma$ controls the spatial spread of the prior.  To match the scale of $\widehat{\mathbf{S}}_j$, we linearly map the heatmap into [-1,1] in order to  balance channel magnitudes and improve training stability\cite{He2015DelvingDI}. ${\mathbf{C}}\in[0,1]$ is then normalized and linearly mapped to $\widehat{\mathbf{C}}$. If auxiliary features are present, they are standardized (z-score) and fused via the lightweight FiLM\cite{perez2018film} modulation inside the denoiser. 
\begin{equation}
\begin{aligned}
\widehat{\mathbf{C}}=2\,\mathbf{C}-1,
\end{aligned}
\end{equation}
where $\widehat{\mathbf{C}}\in[-1,1]^{1\times H\times W}$. Training follows the CCDDPM forward process with a fixed noise schedule $\{\beta_t\}_{t=1}^{T}$. The noisy image $\widehat{\mathbf{S}}_j(t)$ at any time t can be directly obtained by linearly adding the original clean image $\widehat{\mathbf{S}}_j$ and a random noise ${\epsilon}$:
\begin{equation}
\widehat{\mathbf{S}}_j(t)=\sqrt{\bar{\alpha}_t}\,\widehat{\mathbf{S}}_j+\sqrt{1-\bar{\alpha}_t}\,\boldsymbol{\epsilon},\boldsymbol{\epsilon}\sim\mathcal{N}(\mathbf{0},\mathbf{I}),
\end{equation}
\begin{equation}
\bar{\alpha}_t=\prod_{s=1}^{t}(1-\beta_s),\ \ 
\end{equation}
where $\bar{\alpha}_t$ represents the cumulative effect of noise from the beginning to the current step, and $\beta_s$ defines the intensity of noise at each step.

The denoiser is a 2D U-Net $\epsilon_{\boldsymbol{\theta}}$ that predicts the injected noise at a given diffusion step and $\widehat{\boldsymbol{\epsilon}}_t$ is the noise predicted by the model at time step t\cite{10.1007/978-3-319-24574-4_28}. We also difine $\widehat{\mathbf{S}}_j$ as the noisy image state obtained from the original image after t steps of denoising process. Its input is the channel-wise concatenation $[\widehat{\mathbf{S}}_j(t)\oplus\widehat{\mathbf{C}}]\in\mathbb{R}^{2\times H\times W}$ and its output is a single-channel noise field. The pyramid uses downsampling channels with two residual blocks per scale and a mirrored upsampling path with skip connections to preserve high-resolution details. Spatial self-attention\cite{vaswani2017attention} layers are inserted to capture long-range interactions guided by the coordinate prior. SiLU activations with Group/LayerNorm are used throughout. Formally, for time step $t$:
\begin{equation}
\widehat{\boldsymbol{\epsilon}}_t
=\epsilon_{\boldsymbol{\theta}}\!\big([\widehat{\mathbf{S}}_j(t)\oplus\widehat{\mathbf{C}}],\,t\big).
\label{eq:denoiser}
\end{equation}
where $\epsilon_{\boldsymbol{\theta}}$ is the trainable denoising function estimating the noise in the reverse process.



The diffusion index $t$ is encoded by sinusoidal positional embeddings and mapped by a multilayer perceptron (MLP) to obtain a time embedding vector:
\begin{equation}
\mathbf{e}_t \;=\; \mathrm{MLP}(g(t))\in\mathbb{R}^{D},
\label{eq:8}
\end{equation}
where the vector $\mathbf{e}_t$ serves as the global time condition and is injected into residual blocks via FiLM modulation,$D$ is the dimension of the timestep embedding, and $g(\cdot)$ represents the sinusoidal positional embeddings function. In Eq. (9), $\gamma_t$ and $\beta_t$ are channel-wise FiLM scale and shift parameters generated from the timestep embedding $\mathbf{e}_t$, enabling the denoiser to adapt its feature normalization to different noise levels.For a feature map $\mathbf{h}\in\mathbb{R}^{C\times H_l\times W_l}$ inside a residual block, FiLM produces channel-wise scale and shift parameters from $\mathbf{e}_t$ and $l$ denotes the $l$-th resolution level in the U-Net pyramid:
\begin{equation}
(\boldsymbol{\gamma}_t,\boldsymbol{\beta}_t)=\mathbf{W}\mathbf{e}_t+\mathbf{b},
\end{equation}
\begin{equation}
\mathrm{FiLM}(\mathbf{h},t)=(1+\boldsymbol{\gamma}_t)\odot \mathbf{h}+\boldsymbol{\beta}_t,
\tag{10}
\end{equation}
where $\odot$ denotes channel-wise broadcasting. The FiLM conditioning depends on timestep $t$ (through $\mathbf{e}_t$) and does not alter the network topology. Then the encoder down-samples features to enlarge the receptive field.
\begin{equation}
\mathbf{h}_{l+1}=\Psi(\mathbf{h}_l),
\end{equation}
where $\Psi(\cdot)$ represents the down-sampling function,each scale in the pyramid applies residual blocks (two per scale in our implementation) to refine features while preserving stability:
\begin{equation}
\mathbf{h}^{(l,k)} \;=\; \mathbf{h}^{(l,k-1)} \;+\; \mathcal{R}\!\left(\mathbf{h}^{(l,k-1)},\,\mathbf{e}_t\right),
k=1,\dots,K_l,
\tag{11}
\end{equation}
where the mapping function $\mathcal{R}(\cdot)$ denotes a standard Conv--Norm--SiLU stack, where intermediate activations are modulated by $\mathrm{FiLM}(\cdot,t)$ to adapt denoising behavior across noise levels.

Through the upsampling with skip connections, the decoder mirrors the process with skip connections to preserve high-resolution details:
\begin{equation}
\tilde{\mathbf{h}}_l=\Phi(\mathbf{h}_{l+1}),
\end{equation}
\begin{equation}
\mathbf{h}^{\mathrm{dec}}_l=\mathrm{Concat}(\tilde{\mathbf{h}}_l,\mathbf{s}_l),
\end{equation}
where $\Phi(\cdot)$ represents the up-sampling function, $\mathbf{s}_l$ is the encoder feature saved at scale $l$, and $\mathrm{Concat}(\cdot)$ concatenates encoder--decoder features along channels. This mirrored upsampling path with skip connections restores spatial resolution while retaining fine structure.

To capture the long-range spatial dependencies, we insert spatial self-attention at selected intermediate resolutions. 
Flattening the feature map at scale $l$ into $\mathbf{X}\in\mathbb{R}^{(H_lW_l)\times C_l}$, we compute the query, key, and value matrices as linear projections of $\mathbf{X}$, i.e., $\mathbf{Q}$, $\mathbf{K}$, and $\mathbf{V}$ encode what to match, what to be matched against, and what information to aggregate, respectively. 
The attention block is:

\begin{equation}
\mathrm{Attn}(\mathbf{X})=\mathrm{Softmax}\!\left(\frac{\mathbf{Q}\mathbf{K}^\top}{\sqrt{d}}\right)\mathbf{V},
\end{equation}
\begin{equation}
\mathbf{Q}=\mathbf{X}\mathbf{W}^Q,\;
\mathbf{K}=\mathbf{X}\mathbf{W}^K,\;
\mathbf{V}=\mathbf{X}\mathbf{W}^V,
\end{equation}
This self-attention operation is applied in a residual form: $\mathbf{X}\leftarrow \mathbf{X}+\mathrm{Attn}(\mathbf{X})$,which facilitates information exchange between distant spatial locations within the feature map.

At the end, we can formulate the loss function of the CCDDPM model as:
\begin{equation}
\mathcal{L}(\boldsymbol{\theta})=
{E}_{(\widehat{\mathbf{S}}_j,\widehat{\mathbf{C}}),\,t,\,\boldsymbol{\epsilon}}
\!\left[\ \big\|\epsilon_{\boldsymbol{\theta}}([\widehat{\mathbf{S}}_j(t)\oplus\widehat{\mathbf{C}}],t)-\boldsymbol{\epsilon}\big\|_2^2\ \right].
\label{eq:loss}
\end{equation}

As a result, given a target transmitter vehicle coordinate $(x_0,y_0)$, we can construct the corresponding $\ {\mathbf{C}}$ and stack them into $\widehat{\mathbf{C}}\in\mathbb{R}^{1\times H\times W}$. Starting from $\mathbf{S}_{i}(t)$, we run $t$ reverse steps according to the update rule of CCDDPM 
\begin{equation}
\widehat{\mathbf{S}}_{i}(t-1)=\frac{1}{\sqrt{\alpha_t}}\!\left(
\widehat{\mathbf{S}}_t(t)-\frac{1-\alpha_t}{\sqrt{1-\bar{\alpha}_t}}\,
\widehat{\boldsymbol{\epsilon}}_t
\right)
+\sigma_t\,\mathbf{z},
\label{eq:reverse}
\end{equation}
where $\mathbf{z}\!\sim\!\mathcal{N}(\mathbf{0},\mathbf{I})$ for $t>0$. The final sample $\widehat{\mathbf{S}}_0$ is linearly mapped back to 8-bit grayscale to obtain the signal-strength map $\mathbf{S}_0$. For faster sampling,we consider that the Gaussian coordinate prior remains visible to the denoiser across all noise levels, forming a stable spatial anchor at $(x,y)$ while attention layers coordinate far-field decay and shadowing patterns, leading to accurate and condition-consistent REM prediction.

\begin{algorithm}[t]
\small
\caption{CCDDPM with Dynamic Predicted V2X REM Dataset Update}
\label{alg:cond-ddpm}
\textbf{Dataset.} REMs $\mathbf{S_{j}}\in\mathbb{R}^{H\times W}$ is consist of original V2X REM datasets and predicted V2X REM Datasets and each map has a transmitter (Tx).\\
\textbf{Coordinate system.} Map origin at top-left; $x\in[0,255]$ (left$\rightarrow$right), $y\in[0,255]$ (top$\rightarrow$bottom).

\begin{algorithmic}[1]
\State \textbf{Initialize} UNet $\epsilon_\theta$ (input $2\times H\times W$, output $1\times H\times W$), DDPM scheduler ($T{=}1000$), AdamW($\theta$).
\State \textbf{Define} coordinate encoder (Gaussian map)
$c(u,v;x,y)=\exp\big(-((u-x)^2+(v-y)^2)/(2\sigma^2)\big)$;
normalize to $[0,1]$, then map to $[-1,1]$.

\State \textbf{Training iteration} $n=1\dots N$:
\begin{algorithmic}[1]
  \State Sample a mini-batch $\widehat{\mathbf{S}}_j$; scale to $[-1,1]$.
  \State For each $\widehat{\mathbf{S}}_j$, extract Tx $(x,y)$ by argmax intensity (use geometric center if ties).
  \State Build condition $\mathbf{C}\in\mathbb{R}^{1\times H\times W}$ via the Gaussian encoder (with $\sigma\approx5$).
  \State Sample $t\sim\mathcal{U}\{0,\dots,T{-}1\}$ and $\epsilon\sim\mathcal{N}(0,\mathbf{I})$.
  \State Construct noisy input:
  $\widehat{\mathbf{S}}_j(t)=\sqrt{\bar\alpha_t}\,\widehat{\mathbf{S}}_j+\sqrt{1-\bar\alpha_t}\,\epsilon$.
  \State Concatenate model input:  $\texttt{inp}=[\widehat{\mathbf{S}}_j(t),\,\widehat{\mathbf{C}}] \in \mathbb{R}^{2\times H\times W}$.
  \State Predict noise: $\widehat{\boldsymbol{\epsilon}}_t=\epsilon_\theta(\texttt{inp}, t)$.
  \State Compute loss: $\mathcal{L}=\|\widehat{\boldsymbol{\epsilon}}_t-\epsilon\|_2^2$.
  \State Update $\theta$ by AdamW on $\nabla_\theta \mathcal{L}$; apply grad clip and LR schedule.
\end{algorithmic}
\vspace{1mm}
\State Save checkpoint $\theta^\star$.
\vspace{1mm}
\State \textbf{Sampling (inference)} for any coordinate query $(x,y)$:
\begin{algorithmic}[1]
  \State Build condition $\mathbf{C}$ from $(x,y)$ as above.
  \State Initialize $\mathbf{z}\sim\mathcal{N}(0,\mathbf{I})$.
  \For{$t=T,\dots,1$}
    \State $\texttt{inp}=[\widehat{\mathbf{S}}_i(t),\,\widehat{\mathbf{C}}]$; $\widehat{\boldsymbol{\epsilon}}_t=\epsilon_\theta(\texttt{inp}, t)$.
    \State $\widehat{\mathbf{S}}_{i}(t-1)\leftarrow{SchedulerStep}(\widehat{\mathbf{S}}_i(t),\widehat{\boldsymbol{\epsilon}}_t, t)$.
  \EndFor
  \State Map $\mathbf{S}_0$ is from $[-1,1]$ to $[0,255]$ and output a predicted V2X REM. Then predicted V2X REM will be saved to the predicted V2X REM datasets.
\end{algorithmic}
\end{algorithmic}
\end{algorithm}

\subsection{Problem Formulation}
In this paper, we consider a 6G V2X REM and a training set, which is signal–strength maps concatenates the Tx
coordinate with optional environment–auxiliary features. maps are linearly scaled to
$\widehat{\mathbf{S}}_j$; coordinates are encoded into a
continuous Gaussian heatmap $\mathbf{C}$. Our
goal is to learn a conditional generator that, for any query $u=(x,y,\mathrm{env})$,
produces a faithful REM consistent with the scene.

We instantiate with a coordinate–conditioned DDPM.  Let
$\{ \beta_t \}_{t=1}^{T}$ be a fixed noise schedule with
$\alpha_t$ and $\bar\alpha_t$.  The forward process and the $\epsilon$–prediction denoiser $\epsilon_\theta$ induce the following training objective and to ensure the stability of training, the train loss is averaged over multiple randomness factors:
\begin{equation}
\begin{aligned}
\label{eq:cond-ddpm-obj}
 \min_{\theta}\;
& {E}_{(X_j,u)\sim\mathcal{S}}\;
 {E}_{t}\;
 {E}_{\epsilon\sim\mathcal{N}(0,I)}
 \Big[
 \big\|\epsilon_\theta\!\big(\widehat{\mathbf{S}}_j(t),\; C(u),\; t\big)-\epsilon\big\|_2^2
 \Big]\\
& \text{s.t.}\quad
\widehat{\mathbf{S}}_j(t)=\sqrt{\bar\alpha_t}\,\widehat{\mathbf{S}}_j+\sqrt{1-\bar\alpha_t}\,\epsilon.
\end{aligned}
\end{equation}
 
After model training, during the inference phase, we proceed to the inference stage. The learned sampler constructs a mapping from coordinates to signal strengths by iteratively executing reverse diffusion updates; the output is generating REM for the given coordinate $(x_0, y_0)$. To evaluate the consistency between the generated results and the true signal field, we extract the slice at $x = x_0$ and $y = y_0$  to compute its root mean square error (RMSE) following the principle:
\begin{equation}
\mathrm{RMSE}_x(x_0)=
\sqrt{\frac{1}{H}\sum_{y=0}^{H-1}\big(I_{\text{gen}}(y,x_0)-I_{\text{orig}}(y,x_0)\big)^2}.
\label{eq:rmse}
\vspace{-0.5cm}
\end{equation}


\begin{table}[t]
\centering
\caption{Layer specification of CCDDPM.}
\footnotesize
\setlength{\tabcolsep}{3.5pt}
\resizebox{.38\textwidth}{!}{
\begin{tabular}{|p{2.5cm}|p{3.1cm}|p{2.4cm}|}
\hline
\textbf{Type of layer} & \textbf{Kernel / Annotation} & \textbf{Output size} \\
\hline
\multicolumn{3}{|c|}{\textbf{Input \& Conditioning} ($H{=}W{=}256$)}\\ \hline
Input & Concat noisy $x_t$(1ch) + coord C(1ch) & $256\!\times\!256\!\times\!2$ \\ \hline
Coord Encoder & $c(u,v;x,y)\!=\!\exp(-((u-x)^2+(v-y)^2)/(2\sigma^2)),\ \sigma\!=\!5$ & $256\!\times\!256\!\times\!1$ \\ \hline
\multicolumn{3}{|c|}{\textbf{Downsampling path}}\\ \hline
Conv+GN+SiLU & $3\times3$, ch=64 & $256\!\times\!256\!\times\!64$ \\ \hline
Residual Block & $3\times3$, FiLM($t$) & $256\!\times\!256\!\times\!64$ \\ \hline
Downsample & $2\times2$, stride 2 & $128\!\times\!128\!\times\!64$ \\ \hline
Conv+GN+SiLU & $3\times3$, ch=128 & $128\!\times\!128\!\times\!128$ \\ \hline
Residual Block & $3\times3$, FiLM($t$) & $128\!\times\!128\!\times\!128$ \\ \hline
Downsample & $2\times2$, stride 2 & $64\!\times\!64\!\times\!128$ \\ \hline
Conv+GN+SiLU & $3\times3$, ch=256 & $64\!\times\!64\!\times\!256$ \\ \hline
Residual Block & $3\times3$, FiLM($t$) & $64\!\times\!64\!\times\!256$ \\ \hline
Self-Attention & MHSA @ $64\!\times\!64$ & $64\!\times\!64\!\times\!256$ \\ \hline
\multicolumn{3}{|c|}{\textbf{Bottleneck}}\\ \hline
ResBlk $\to$ Attn $\to$ ResBlk & $3\times3$ + MHSA + $3\times3$ & $32\!\times\!32\!\times\!256$ \\ \hline
\multicolumn{3}{|c|}{\textbf{Upsampling path}}\\ \hline
Upsample & $\times2$ & $64\!\times\!64\!\times\!256$ \\ \hline
Skip Concat & from Down4 & $64\!\times\!64\!\times\!256$ \\ \hline
Residual Block $\times2$ & $3\times3$, FiLM($t$) & $64\!\times\!64\!\times\!256$ \\ \hline
Upsample & $\times2$ & $128\!\times\!128\!\times\!256$ \\ \hline
Residual Block $\times2$ & $3\times3$, FiLM($t$) & $128\!\times\!128\!\times\!256$ \\ \hline
Upsample & $\times2$ & $256\!\times\!256\!\times\!128$ \\ \hline
Residual Block $\times2$ & $3\times3$, FiLM($t$) & $256\!\times\!256\!\times\!64$ \\ \hline
\multicolumn{3}{|c|}{\textbf{Output Head}}\\ \hline
Conv+GN+SiLU & $3\times3$ & $256\!\times\!256\!\times\!64$ \\ \hline
Conv (1 ch) & $1\times1$ & $256\!\times\!256\!\times1$ \\ \hline
\multicolumn{3}{|c|}{\textbf{Embeddings / Schedulers}}\\ \hline
Time Embedding & sinusoidal $\!t$-embed $\!\to$ MLP(256), FiLM scale/shift & -- \\ \hline
Sampler (inf.) & CCDDPM $S\!=\!50$ (DDIM/DPMSolver opt.) & -- \\ \hline
\end{tabular}}
\label{tab:cond-diff-layers-single}
\vspace{-0.5cm}
\end{table}

\subsection{Training CCDDPM}
The Model parameters for conditional diffusion is summarized in Table I. The training procedure of the proposed CCDDPM is summarized in Algorithm~\ref{alg:cond-ddpm}. In each iteration, only the denoiser parameters $\theta$ are updated while the diffusion schedule is kept fixed. Given REMs and their transmitter coordinates, we (i) build the continuous coordinate prior by encoding $(x,y)$ into a Gaussian heatmap and concatenate it with the noisy image channel; (ii) draw a timestep $t$ and Gaussian noise , and synthesize the latent ${\mathbf{S}}_j(t)$ via the forward process; (iii) feed $[\widehat{\mathbf{S}}_j(t)\oplus\widehat{\mathbf{C}}]$ and $t$ into the conditional U-Net $\epsilon_\theta$ to predict the injected noise; and (iv) update $\theta$ by minimizing the standard $\epsilon$-prediction objective using AdamW with cosine decay, warm-up, mixed precision, and gradient-norm clipping for stability. Checkpoints are saved periodically and the best model is selected by validation loss.At inference, the learned denoiser is frozen. For any query coordinate $(x,y)$, we construct its heatmap $\mathbf{C}$ once, sample $S_i(t)$, and run $t$ reverse steps of the CCDDPM sampler to obtain $\widehat{S}_i(t)$ while conditioning on $\mathbf{C}$ at every step. The final result is linearly mapped from $[-1,1]$ back to $[\mathbf{S}_j^{min},\mathbf{S}_j^{max}]$ to yield the REM for the given coordinate. 



\section{Experimental Evaluation}
\label{sec:sim}

Performance evaluation is conducted on Google Colab with Python 3 Google Compute Engine backend (GPU) comprising 220 compute units.
The \textbf{Algorithm 1} in Section 2 was implemented in Python using PyTorch \footnote{All code used to generate the results can be found
at \url{https://github.com/lzy1207/v2x_signal} }. \textbf{We use 1000 V2X REMs with $256{\times}256$ resolution(meters) for training}, Unless otherwise specified, we train on 900 REMs and evaluate on the rest 100 REMs. 



\begin{figure}[t]
    \centering
    \begin{minipage}[t]{0.22\textwidth}
        \centering
\includegraphics[width=\linewidth]{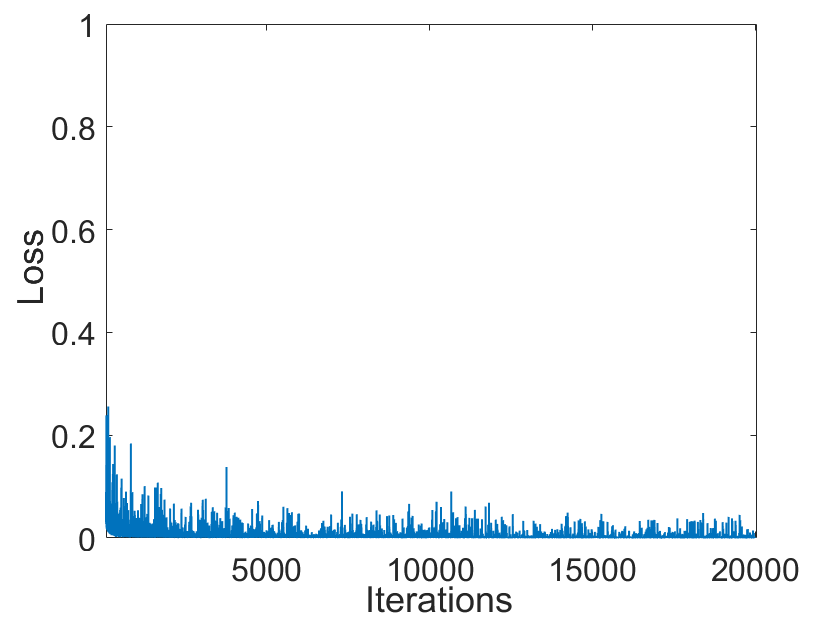}
        \caption{Training Loss of CCDDPM.}
                \vspace{-0.6cm}
        \label{fig:train_loss}
    \end{minipage}
    \hfill 
    \begin{minipage}[t]{0.26\textwidth}
        \centering
\includegraphics[width=\linewidth]{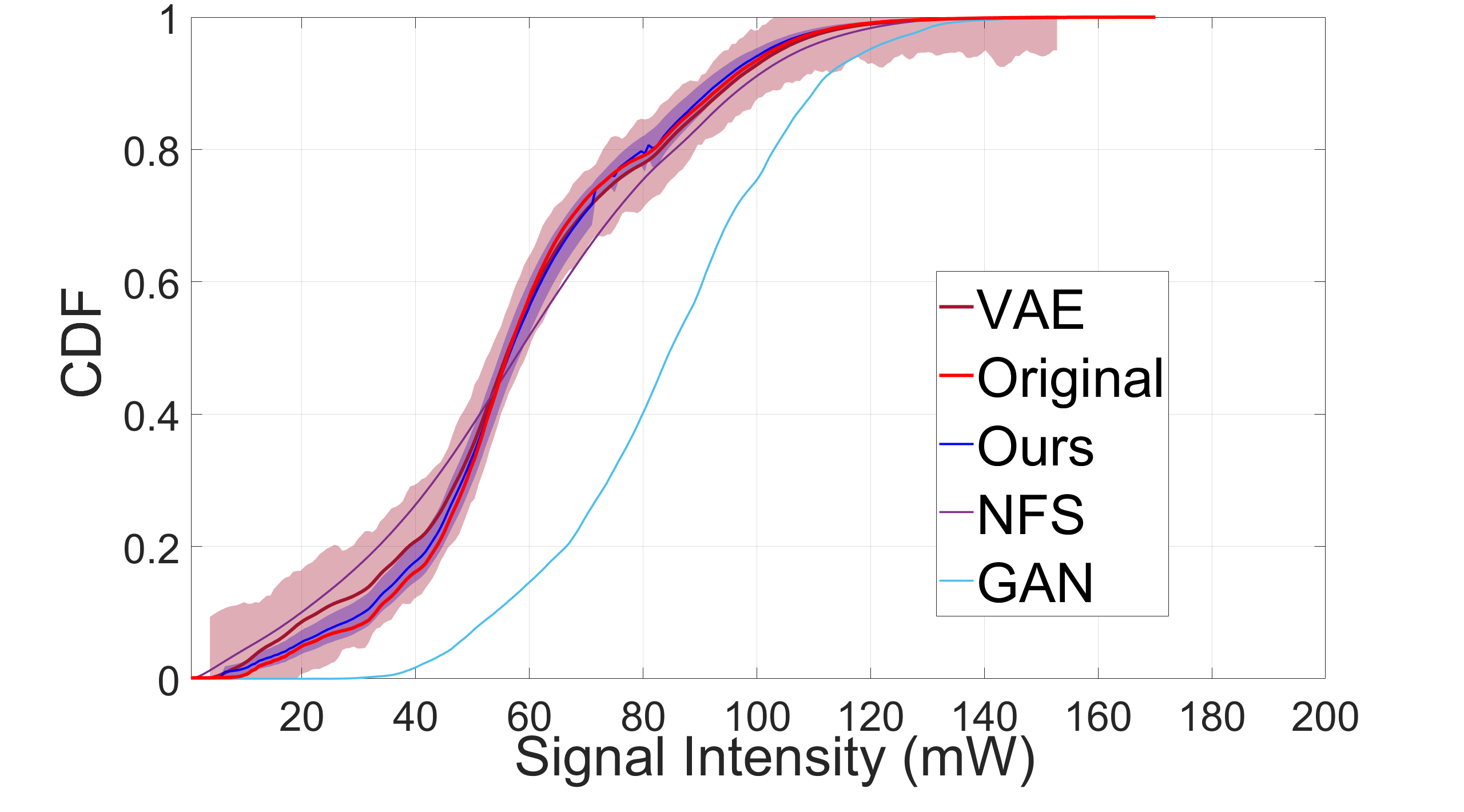}
        \caption{Signal Intensity CDF at $(x_0,y_0)=(108,178)$.}
        \vspace{-0.6cm}
        \label{fig:CDF}
    \end{minipage}
\end{figure}

As shown in Fig.~\ref{fig:train_loss}, we train the proposed CCDDPM with 20000 iterations. The loss decreases overall with occasional spikes, which is typical for diffusion training where timesteps and Gaussian noise are sampled on-the-fly. It drops sharply within the first 2000 iterations, then continues to decline with small oscillations. A stable region appears around 12000 iterations, where the loss stays below 0.02, and by 20000 iterations it converges to 0.01 with only minor batch-level fluctuations. The intermittent peaks likely come from batches with large timesteps (higher noise) or scenarios with strong occlusions that make denoising harder. 

Qualitative and quantitative comparison based on a random transmitter coordinator $(x_0,y_0)=(108,178)$. We first compare the signal intensity CDFs of the corresponding REM generated by different generative approaches \cite{van2024generative}.  As shown in Fig.~\ref{fig:CDF}, the CDF of our CCDDPM almost overlaps with the original curve across the entire range of signal intensity, achieving the best distributional consistency. In contrast, Normalizing Flows (NFs), Generative Adversarial Network	(GAN), and Variational Autoencoder (VAE) show varying deviations; VAE is relatively closer to the original REM, while the other two deviate more in the mid-to-high intensity region. To assess stability, we independently generate 100 REMs at the same coordinate with CCDDPM and VAE, and draw the mean and the standard deviation envelopes at each intensity. Our CCDDPM yields a smaller error region indicating lower sampling variance and better repeatability, and its mean curve remains closest to the Original. Hence, CCDDPM outperforms the other generative methods in both distributional fidelity and generation stability.




\begin{figure*}[!t]
\begin{minipage}[t]{0.18\linewidth}
\centering
 \subfigure[Original.]{
\includegraphics[width=1.0in]{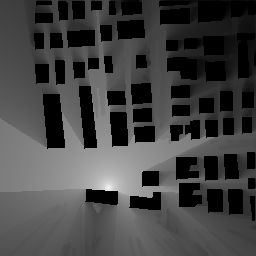}}
\label{fig:3a}
\end{minipage}%
\begin{minipage}[t]{0.18\linewidth}
\centering
 \subfigure[Ours.]
{\includegraphics[width=1.0in]{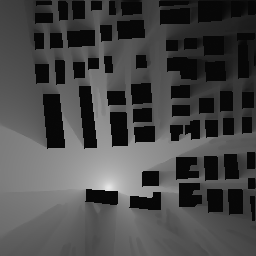}}
\label{fig:3b}
\end{minipage}
\begin{minipage}[t]{0.18\linewidth}
\centering
\subfigure[VAE.]{
\includegraphics[width=1.0in]{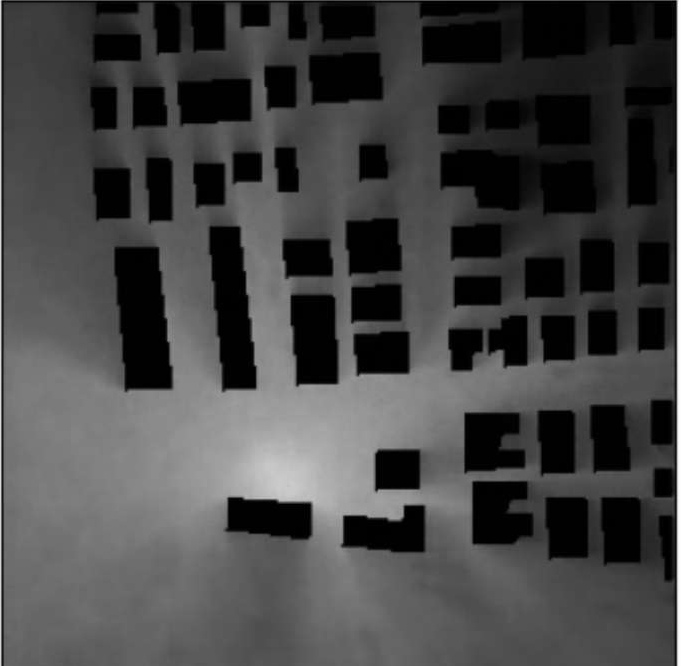}}
\label{fig:3c}
\end{minipage}
\begin{minipage}[t]{0.18\linewidth}
\centering
\subfigure[GAN.]{
\includegraphics[width=1.0in]{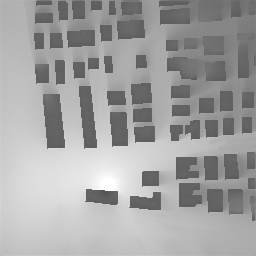}}
\label{fig:3d}
\end{minipage}
\begin{minipage}[t]{0.18\linewidth}
\centering
\subfigure[NFs.]{
\includegraphics[width=1.0in]{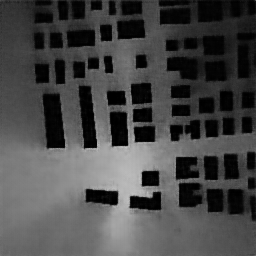}}
\label{fig:3e}
\end{minipage}
\vspace{-0.2cm}
\caption{Signal Intensity-based REM Comparison Among Different Generative Approaches.}
\vspace{-0.5cm}
\label{fig:qual_comp}
\end{figure*}

 As Fig. \ref{fig:qual_comp} shows, we further compare the original REM with generative ones. Our CCDDPM best preserves the near-field hotspot around the Transmitter and the elongated shadowing lobes cast by surrounding blocks; building boundaries remain sharp while the long-range decay is smooth and scene-consistent. In contrast, VAE tends to over-smooth high-gain regions,  GAN shows mild ringing and misplaced bright ridges, and NFs exhibit contrast drift with reduced dynamic range. Overall, CCDDPM produces the closest result to the ground truth among all baselines. 
 
\begin{figure}[t]
\begin{minipage}[t]{0.48\linewidth}
\centering
 \subfigure[Scenario 1: $x=100$.]
{\includegraphics[width=1.8in]{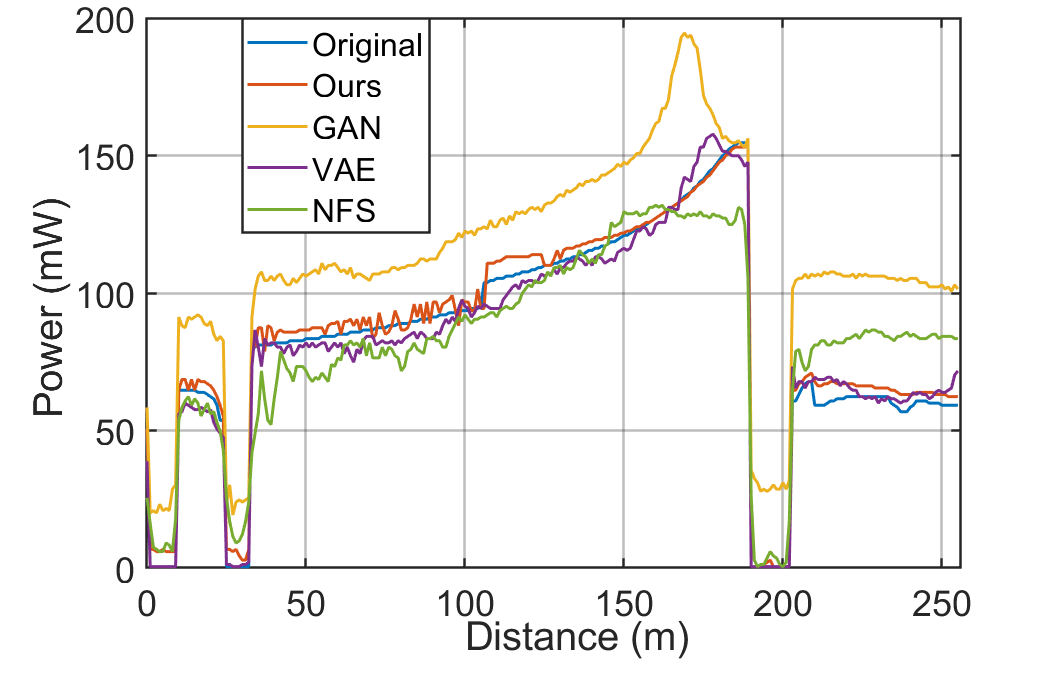}}
\label{fig:4a}
\end{minipage}
\begin{minipage}[t]{0.24\linewidth}
\centering
\subfigure[Scenario 2: $y=100$.]{
\includegraphics[width=1.7in]{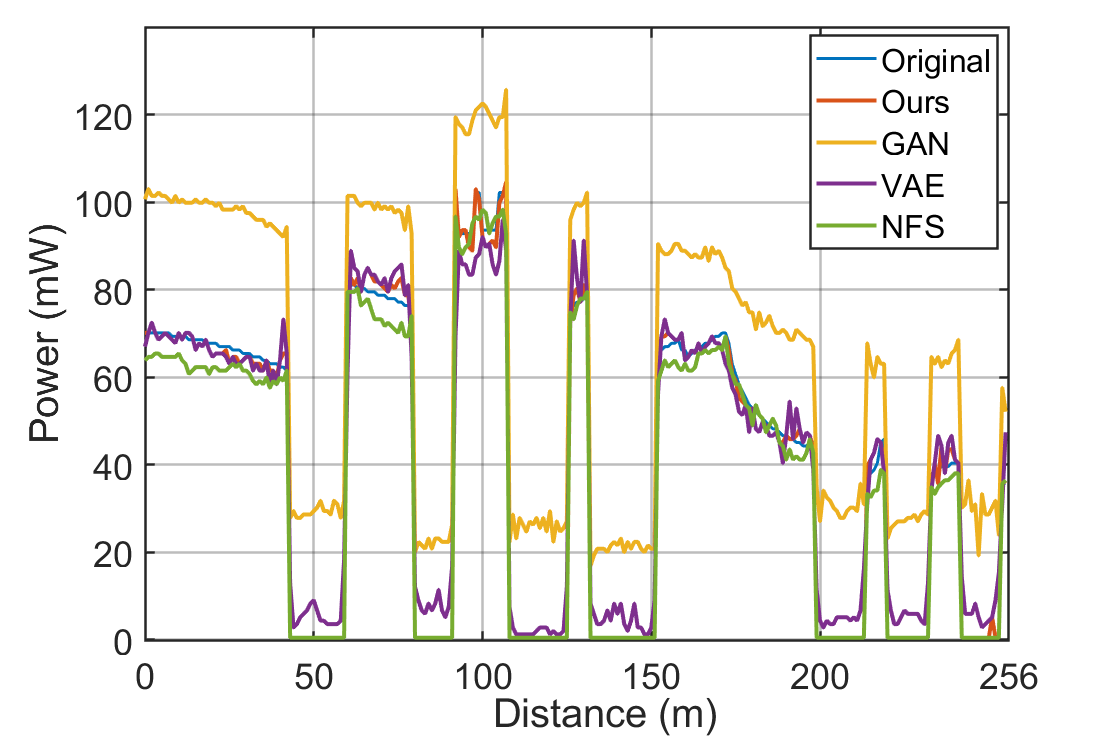}}
\label{fig:4b}
\end{minipage}%
\vspace{-0.2cm}
\caption{RMSE Comparison.}
\vspace{-0.6cm}
\label{fig:RMSE}
\end{figure}

 In Fig.~\ref{fig:RMSE}, we compare the line-slice RMSE between each generated REM and the original REM to evaluate the robustness. To account for sampling randomness, we generate multiple realizations at the same coordinate, randomly select one per method, and compute its RMSE. Both scenarios ((vertical slice at $x=100$ and horizontal slice at $y=100$))
 show the same trend where our CCDDPM remains closest to the original one, whereas NFs and conditional GAN incur larger RMSE, particularly around occlusion boundaries with rapid intensity changes. Thus, our CCDDPM achieves the lowest average RMSE and the smallest variability.

\section{Conclusion}
\label{sec:con}
In this paper, we develop the CCDDPM that is a diffusion-based generative approach to predict the high-fidelity signal intensity-based REM towards 6G V2X according to the learned distribution of the historical and predicted V2X REMs. REMs predicted by the CCDDPM closely follow ground-truth statistics and spatial structure of the signal intensity-based REM datasets. Our evaluation shows the CCDDPM performs higher fidelity and stability over other generative approaches. The resulting predictor can rapidly generate scenario-consistent REMs for arbitrary transmitter coordinates, thereby enabling more efficient 6G C-V2X communications by reducing the likelihood that transmitter vehicles experience PHY-layer impairments.



\bibliographystyle{IEEEtran}
\bibliography{reference}

@article{yapar2022dataset,
  title={Dataset of pathloss and ToA radio maps with localization application},
  author={Yapar, {\c{C}}a{\u{g}}kan and Levie, Ron and Kutyniok, Gitta and Caire, Giuseppe},
  journal={arXiv preprint arXiv:2212.11777},
  year={2022}
}

@ARTICLE{10440080,
  author={Khan, Shafi Ullah and García, Carla E. and Hwang, Taewoong and Koo, Insoo},
  journal={IEEE Access}, 
  title={Radio Environment Map Construction Based on Privacy-Centric Federated Learning}, 
  year={2024},
  volume={12},
  number={},
  pages={28109-28121},
  doi={10.1109/ACCESS.2024.3367589}}

@INPROCEEDINGS{10333693,
  author={Lu, Yang and Zhang, Yifan and Shi, Tuo and Wang, Jianping and Wu, Jen-Ming and Liu, Bingyi},
  booktitle={2023 IEEE 98th Vehicular Technology Conference (VTC2023-Fall)}, 
  title={Empirical Study and Signal Intensity Prediction for Cellular Vehicle-to-Everything (C-V2X)}, 
  year={2023},
  volume={},
  number={},
  pages={1-6},
  doi={10.1109/VTC2023-Fall60731.2023.10333693}}

@INPROCEEDINGS{10437765,
  author={Mo, Ronghong and Pei, Yiyang and Sun, Sumei and Premkumar, A. B. and Venkatarayalu, Neelakantam V},
  booktitle={GLOBECOM 2023 - 2023 IEEE Global Communications Conference}, 
  title={Federated Learning-Based Radio Environment Map Construction for Wireless Networks}, 
  year={2023},
  volume={},
  number={},
  pages={5238-5243},
  doi={10.1109/GLOBECOM54140.2023.10437765}}

@ARTICLE{10554663,
  author={Chaccour, Christina and Saad, Walid and Debbah, Mérouane and Han, Zhu and Vincent Poor, H.},
  journal={IEEE Communications Surveys and Tutorials}, 
  title={Less Data, More Knowledge: Building Next-Generation Semantic Communication Networks}, 
  year={2025},
  volume={27},
  number={1},
  pages={37-76},
  doi={10.1109/COMST.2024.3412852}}

@INPROCEEDINGS{10898772,
  author={Indulekha, K P and Venkatesh, T. G.},
  booktitle={2024 IEEE International Conference on Advanced Networks and Telecommunications Systems (ANTS)}, 
  title={{Dynamic Resource and Power Allocation strategy for ISAC-Aided 5G V2X Networks using Reinforcement learning}}, 
  year={2024},
  volume={},
  number={},
  pages={1-6},
  doi={10.1109/ANTS63515.2024.10898772}}

@article{van2024generative,
  title={Generative AI for physical layer communications: A survey},
  author={Van Huynh, Nguyen and Wang, Jiacheng and Du, Hongyang and Hoang, Dinh Thai and Niyato, Dusit and Nguyen, Diep N and Kim, Dong In and Letaief, Khaled B},
  journal={IEEE Transactions on Cognitive Communications and Networking},
  year={2024},
  publisher={IEEE}
}

@article{He2015DelvingDI,
  title={Delving Deep into Rectifiers: Surpassing Human-Level Performance on ImageNet Classification},
  author={Kaiming He and X. Zhang and Shaoqing Ren and Jian Sun},
  journal={2015 IEEE International Conference on Computer Vision (ICCV)},
  year={2015},
  pages={1026-1034},
}

@InProceedings{10.1007/978-3-319-24574-4_28,
author="Ronneberger, Olaf
and Fischer, Philipp
and Brox, Thomas",
editor="Navab, Nassir
and Hornegger, Joachim
and Wells, William M.
and Frangi, Alejandro F.",
title="U-Net: Convolutional Networks for Biomedical Image Segmentation",
booktitle="Medical Image Computing and Computer-Assisted Intervention -- MICCAI 2015",
year="2015",
publisher="Springer International Publishing",
address="Cham",
pages="234--241",
isbn="978-3-319-24574-4"
}

@inproceedings{vaswani2017attention,
title={Attention Is All You Need},
author={Vaswani, Ashish and Shazeer, Noam and Parmar, Niki and Uszkoreit, Jakob and Jones, Llion and Gomez, Aidan N. and Kaiser, Lukasz and Polosukhin, Illia},
booktitle={Advances in Neural Information Processing Systems (NeurIPS)},
pages={5998--6008},
year={2017}
}

@InProceedings{perez2018film,
  title={FiLM: Visual Reasoning with a General Conditioning Layer},
  author={Ethan Perez and Florian Strub and Harm de Vries and Vincent Dumoulin and Aaron C. Courville},
  booktitle={AAAI},
  year={2018}
}

@article{luo2025denoising,
  title={Denoising diffusion probabilistic model for radio map estimation in generative wireless networks},
  author={Luo, Xuanhao and Li, Zhizhen and Peng, Zhiyuan and Chen, Mingzhe and Liu, Yuchen},
  journal={IEEE Transactions on Cognitive Communications and Networking},
  year={2025},
  publisher={IEEE}
}

@article{cao2022resource,
  title={Resource allocation in {5G} platoon communication: Modeling, analysis and optimization},
  author={Cao, Liu and Roy, Sumit and Yin, Hao},
  journal={IEEE Transactions on Vehicular Technology},
  volume={72},
  number={4},
  pages={5035--5048},
  year={2022},
  publisher={IEEE}
}


\end{document}